\documentclass[twocolumn,english,aps, prl,nofootinbib,  superscriptaddress, preprintnumbers,floatfix]{revtex4-2}
\usepackage[T1]{fontenc}
\usepackage{babel}
\usepackage{mathrsfs}
\usepackage{bm,amsmath}
\usepackage{amssymb}

\usepackage{graphicx}
\usepackage{hyperref}
\hypersetup{pdftex,colorlinks=true,linkcolor=blue,citecolor=blue,menucolor=black,urlcolor=blue,filecolor=blue}

\usepackage{soul}
\setstcolor{red}\setul{0.5ex}{0.3ex}

\usepackage{slashed}
\usepackage{gensymb}
\usepackage{bbold}
\usepackage{multirow}

\usepackage{amsfonts}
\usepackage{makecell}
\usepackage[dvipsnames]{xcolor}
\usepackage[normalem]{ulem}
\usepackage{longtable}
\usepackage{array}
\usepackage{float}
\usepackage{subfigure}
\usepackage{orcidlink}

\newcommand{\jpsi}{J/\psi}
\newcommand{\psip}{\psi(2S)}
\newcommand{\phnn}{\Gamma_{\gamma Y Y}}
\newcommand{\phnd}{\Gamma_{\gamma Y^* Y}}
\newcommand{\cnn}{\Gamma_{\jpsi Y\bar{Y}}}
\newcommand{\cnd}{\Gamma_{\jpsi Y^*\bar{Y}}}
\newcommand{\bes}{BES\Rom{3}}

\newcommand*{\Rom}[1]{\uppercase\expandafter{\romannumeral #1\relax}}

\newcommand{\itp}{\affiliation{CAS Key Laboratory of Theoretical Physics, Institute of Theoretical Physics,\\
Chinese Academy of Sciences, Beijing 100190, China}}

\newcommand{\ucas}{\affiliation{School of Physical Sciences, University of Chinese Academy of Sciences, Beijing 100049, China}}

\newcommand{\peng}{\affiliation{Peng Huanwu Collaborative Center for Research and Education, Beihang University, Beijing 100191, China}}

\newcommand{\hiskp}{\affiliation{Helmholtz-Institut f\"ur Strahlen- und Kernphysik and Bethe Center for Theoretical Physics,\\ Universit\"at Bonn, D-53115 Bonn, Germany}}

\newcommand{\fzj}{\affiliation{Institute for Advanced Simulation,  Forschungszentrum J\"ulich, D-52425 J\"ulich, Germany}}

\newcommand{\tbilisi}{\affiliation{Tbilisi State University, 0186 Tbilisi, Georgia}}

\begin{document}
\title{Method for measuring the {charge radii of charged hyperons} from the time-like region}

\author{Yong-Hui Lin\orcidlink{0000-0001-8800-9437}}\email{yonghui@hiskp.uni-bonn.de}\hiskp

\author{Feng-Kun Guo\orcidlink{0000-0002-2919-2064}}\email{fkguo@itp.ac.cn}\itp\ucas\peng

\author{Ulf-G.~Mei{\ss}ner\orcidlink{0000-0003-1254-442X}}\email{meissner@hiskp.uni-bonn.de}\hiskp\fzj\tbilisi

\begin{abstract}
	We propose a novel method for measuring the charge radii of charged stable hadrons, with which the first measurement of the charge radii of the $\Sigma^+$ and the $\Xi^-$ is foreseen. The method explores the facts that the Dalitz decay $\psi(2S) \to Y\bar{Y}e^+e^-$ contains the hyperon form factors and the lowest measurable four-momentum transfer squared  can be as low as $\sim 4m_e^2= 1.05\times10^{-6}\,{\rm GeV^2}$ in the time-like region.
	We identify a kinematic region where the hyperon form factors are essential and propose a method for subtracting the background from the data. 
	It is estimated that the hyperon charge radii can be measured to a precision of about {0.2~fm} with the BES\Rom{3} experiment and one order of magnitude better at the future Super $\tau$-Charm Facility. 
    Moreover, the same method can be used to measure the charge radius of the proton, which provides an independent cross-check on the extraction of proton radius from elastic $ep$ scattering or leptonic hydrogen spectroscopy.
\end{abstract}
\maketitle

\section{Introduction}

The size of a hadron, a consequence of the confinement of quarks and gluons inside a finite volume, is one of the most fundamental and simplest static properties. It is, however, not an unambiguously
defined observable but depends on the probe used in experiments. The electric charge radius of a hadron defines its size seen by a photon probe and is encoded in its electromagnetic form factors (EMFFs). Systematic investigations of the hadron EMFFs from both theoretical and experimental perspectives have led to a better understanding of the internal structure of hadrons and the nonperturbative nature of quantum chromodynamics (QCD) at low energies. For recent reviews, see, e.g., Refs.~\cite{Punjabi:2015bba,Horn:2016rip,Ramalho:2023hqd}. 

Over the past decade, a remarkable surge in research dedicated to the nucleon EMFFs has been witnessed, which was spurred by the report of the first extraction of the proton charge radius from the spectroscopic measurement of muonic hydrogen with unprecedented precision in 2010~\cite{Pohl:2010zza}. This value, $0.84184(67)\,{\rm fm}$~\cite{Pohl:2010zza,Antognini:2013txn}, unexpectedly exhibited a $5\sigma$ deviation from the previously accepted one, $0.8768(69)\,{\rm fm}$~\cite{Mohr:2008fa}, which was based on electron scattering and ordinary hydrogen spectroscopy measurements. Recently, a high-precision analysis, based on dispersion relations, of the world data of the nucleon EMFFs in the space- and time-like regions led to $r^p_E=0.840(3)(4)\,{\rm fm}$~\cite{Lin:2021xrc}, which fully agrees with the latest CODATA value,$0.8414(19)\,{\rm fm}$~\cite{Tiesinga:2021myr}. 
An important ingredient in this analysis is the isovector spectral function~\cite{Hohler:1974eq,Hoferichter:2016duk}. For reviews on the progress on the determination of the proton charge radius, we refer to~\cite{Hammer:2019uab,Karr:2020wgh,Gao:2021sml,Lin:2021umz,Peset:2021iul,Antognini:2022xoo}.

It is important to highlight that the definition of the charge radius, $\langle (r^p_E)^2 \rangle= -6\, d G_E^p/d Q^2|_{Q^2=0}$, where $G_E^p$ is the  electric Sachs form factor of the proton, implies a greater sensitivity to the data at lower four-momentum transfer squared. 
This underscores the significance of  measurements conducted at low momentum transfers in accurately determining the charge radius. Currently, the lowest accessible value of the four-momentum transfer squared for the proton charge radius extraction is reached by the PRad experiment, $Q^2= 2.1\times 10^{-4}$~GeV$^2$\cite{Xiong:2019umf}. The follow-up program PRad-\Rom{2} under construction aims to reach values of $Q^2$ as low as $10^{-5}~{\rm GeV^2}$~\cite{PRad:2020oor}.

While notable progress towards measuring the proton charge radius has been made in recent years, the experimental knowledge of the electromagnetic structure of other baryons and, in particular, of those with strangeness  (the hyperons) is scarce due to absence of stable hyperon targets. So far, only the charge radius of the $\Sigma^-$, among all hyperons, has been measured to be $0.78(10)$~fm using a $\Sigma^-$ beam at a mean energy of 610~GeV~\cite{SELEX:2001fbx}. The fact that the kaon charge radius {{(0.560$\pm 0.031$~fm)}} has been found to be smaller than that of the pion {{(0.659$\pm 0.004$~fm)}}~\cite{ParticleDataGroup:2022pth} suggests a flavor dependence of the hadron size~\cite{Bernard:1988bx,Maris:2000sk,SELEX:2001fbx,Horn:2016rip}.
There is also a large spread of theoretical predictions for the hyperon charge radii, covering the range from 0.6 to 1.0~fm for the $\Sigma^+$ and from 0.4 to 0.7~fm for $\Xi^-$~\cite{Wagner:1998fi,Kubis:2000aa,HillerBlin:2017syu,Yang:2020rpi,Yan:2023yff}.
To clarify how the hadron structure depends on the valence quark content and how the flavor SU(3) symmetry of QCD is broken, a systematic study of the charge radii of hyperons is necessary. 
{Measurements of the hyperon and hyper-nucleus charge radii can also shed light on the baryon-baryon interactions; one particular example of this point is provided by the precise determination of $\langle (r^n_E)^2 \rangle$ from the measurement of $\langle (r^d_E)^2 \rangle-\langle (r^p_E)^2 \rangle$ using the state-of-the-art nucleon-nucleon potentials~\cite{Filin:2019eoe}, where the superscripts $n$ and $d$ denote neutron and deuteron, respectively.}
Additional interest on hyperons comes from the growing evidence that strange particles can have significant implications for astrophysics, such as the reduction of the maximum value of neutron star mass by the hyperon degrees of freedom~\cite{Chatterjee:2015pua,Ramalho:2023hqd}.

In this Letter, we propose a novel method to determine the charge radii of charged hyperons, namely, to measure it in the time-like region in the four-body Dalitz decay $\psi(2S) \to Y\bar{Y}e^+e^-$, making use of the huge $\psi(2S)$ data sets, $2.7\times10^{9}$, collected at BESIII~\cite{BESIII:2024lks} and $6.4\times10^{11}$ at the potential Super $\tau$-Charm Facility (STCF)~\cite{Achasov:2023gey}. Here, $Y$ denotes the $\Sigma^+$ and the $\Xi^-$. It opens an opportunity to access the hyperon EMFFs experimentally at an extremely low four-momentum transfer squared, i.e., $q^2\equiv -Q^2\sim 4m_e^2 = 1.05\times10^{-6}\,{\rm GeV^2}$, since the electron and positron can be detected with a high efficiency as long as their transverse momenta are larger than a few tens of MeV (about 50~MeV at BESIII). 
The low $q^2\simeq4m_e^2$ can be reached by detecting the electron and positron produced with a small relative momentum.
The resulting determination of the charge radii of charged hyperons will exhibit a high accuracy once sufficient events are collected. 
Furthermore, similar reactions with $Y\bar{Y}$ replaced by $p\bar p$ can be used to measure the proton charge radius, which provides an independent cross-check on the extraction of proton radius from elastic $ep$ scattering or leptonic hydrogen spectroscopy, {{although the precision of this novel approach may not be competitive with that of existing high-accuracy measurements, such as the PRad experiment in $ep$ scattering and the CREMA experiment in muonic hydrogen spectroscopy}}.

\section{Dalitz decay \texorpdfstring{$\psi(2S) \to Y\bar{Y}e^+e^-$}{psi2S to Y Ybar e+ e-}}

\begin{figure}[t]
	\centering
	\includegraphics*[width=0.48\textwidth,angle=0]{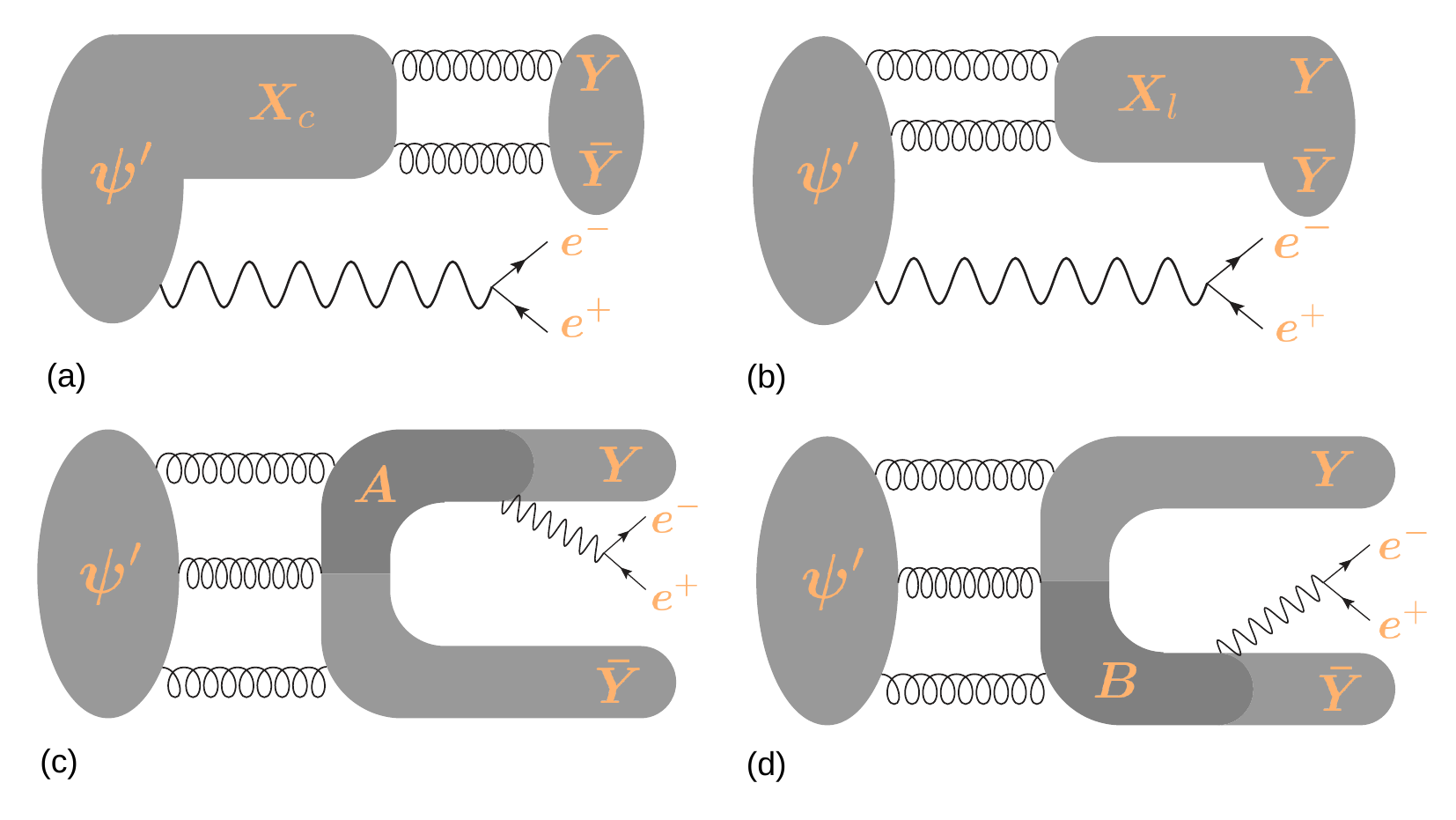}
	\caption{
		Diagrams for the $\psi(2S) \to Y\bar{Y}e^+e^-$ process. The diagrams (a) and (b) are denoted as type-X, for which the $Y\bar Y$ pair is produced via two gluons, (c) as type-A and (d) as type-B, for which the light hadrons are produced via three gluons (or one virtual photon). 
	}
	\label{fig:diagram}
\end{figure}

The considered Dalitz decay can happen through several mechanisms shown in Fig.~\ref{fig:diagram}, following the notation of Ref.~\cite{Kappert:2022fox}: 
\begin{itemize}
	\item type-X: $\psip\to\gamma^* X(X\to Y\bar{Y})$, see diagrams (a) and (b);
	\item type-A: $\psip\to \bar{Y} A(A\to \gamma^* Y)$, see diagram (c);
	\item type-B: $\psip\to Y B(B\to \gamma^* \bar{Y})$, see diagram (d).
\end{itemize}
In diagrams (a) and (b), the virtual photon is emitted from the charm or anti-charm quark in the $\psip$, and the $Y\bar Y$ pair is then mainly produced through two gluons.
In diagrams (c) and (d), the $c\bar c$ pair first annihilates into three gluons (or one virtual photon) that hadronize into a hyperon-antihyperon pair, the virtual photon, which converts to $e^+e^-$, is emitted from the antihyperon or hyperon, respectively, via the final-state radiation (FSR), $\psip \to Y\bar{Y}\gamma_{\rm FSR}^*(\gamma_{\rm FSR}^*\to e^+e^-)$. 
The hyperon EMFFs are embodied in the decay amplitude of the type-A and type-B diagrams. 

\begin{figure}[t]
	\centering
	\includegraphics*[width=0.48\textwidth,angle=0]{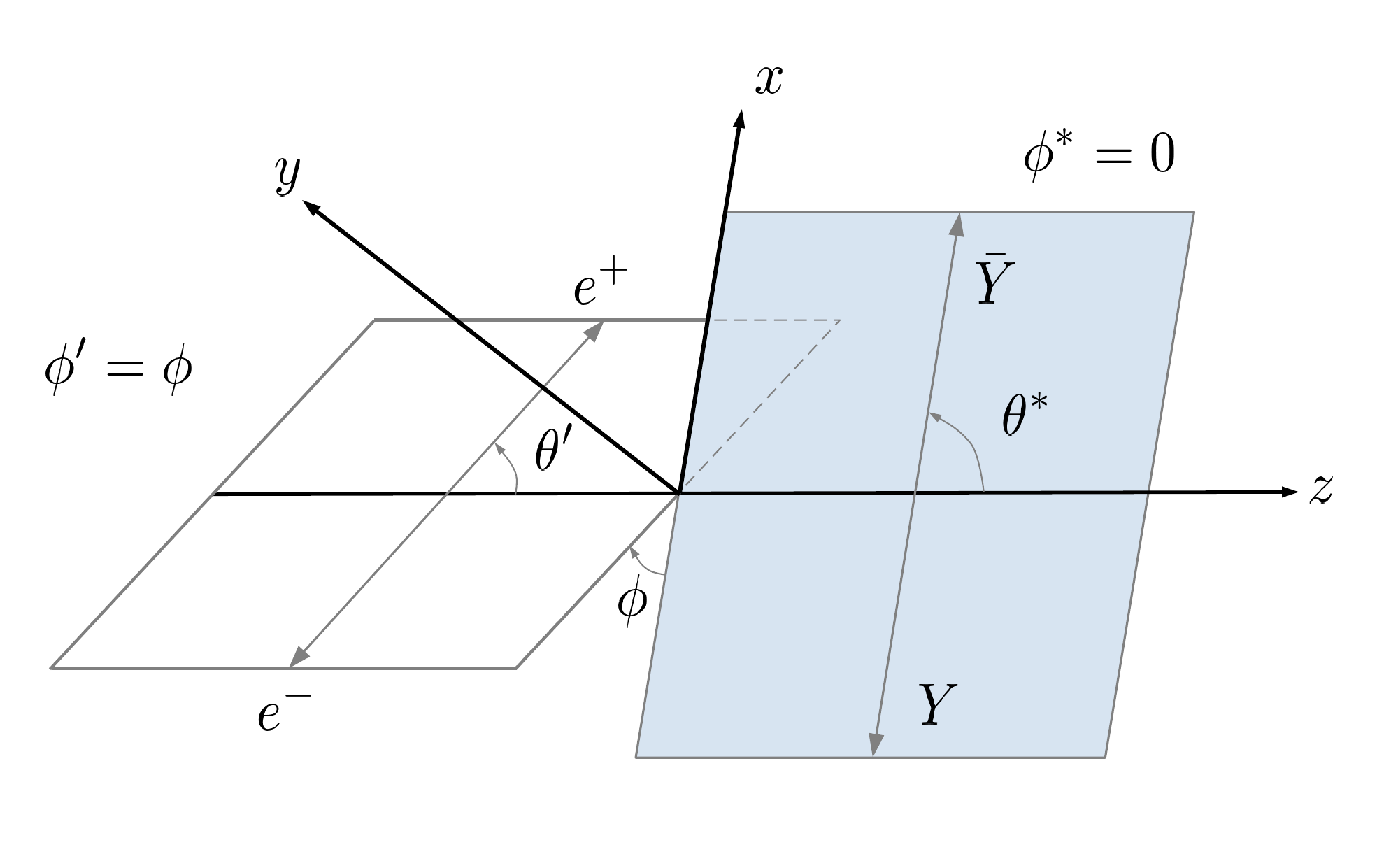}
	\caption{
		Kinematics for the $\psi(2S) \to Y\bar{Y}e^+e^-$ process.
	}
	\label{fig:kinematic}
\end{figure}

The differential rate of the decay process $\psip \to Y\bar{Y}e^+e^-$ can be written as
\begin{align}\label{eq:dcs}
	&\frac{d\Gamma}{d m_{e^+e^-} d m_{Y\bar{Y}}  d\cos\theta^*d\cos\theta^\prime d\phi}\notag\\
	= & \frac{|\vec{k}_{e^+e^-}||\vec{k}_Y^*||\vec{k}_{e^-}^\prime|}{(2\pi)^6 16 M_{\psip}^2}  \frac{C(q^2)}3\sum_{\rm spins}|{\cal M}|^2,
\end{align}
where $\vec{k}_{e^+e^-}$, ${\vec{k}_Y^*}$ and ${\vec{k}_{e^-}^\prime}$ are the three-momentum of $e^+e^-$ in the $\psip$ rest frame, that of $Y$ in the $Y\bar Y$ center-of-mass (c.m.) frame, and that of $e^-$ in the $e^+e^-$ c.m. frame, respectively. {{($\theta^*$, $\phi^*$) and ($\theta^\prime$, $\phi^\prime$) parameterize the directions of three-momentum ${\vec{k}_Y^*}$ and ${\vec{k}_{e^-}^\prime}$ in their respective center-of-mass frames. For the process under study, only one independent azimuthal angle is required, as illustrated in Fig.~\ref{fig:kinematic}, and we adopt the convention $\phi^\prime=\phi$ and $\phi^*=0$.}}
$C(q^2)\equiv y/(1-e^{-y})$ is the Sommerfeld-Gamow factor with $y\equiv\pi\alpha m_e/|\vec{k}_{e^-}^\prime|$ and $\alpha$ the 
fine-structure constant, and
\begin{align}\label{eq:amplitude}
   \left\lvert {\cal M} \right\rvert^2&=\left\lvert {\cal M}_{\rm signal} \right\rvert^2 +\left\lvert {\cal M}_X \right\rvert^2, \\
	\left\lvert {\cal M}_{\rm signal} \right\rvert^2 &\equiv \left\lvert {\cal M}_{A+B} \right\rvert^2 +
 2\operatorname{Re}\left({\cal M}_{A+B}^{}{\cal M}_X^*\right) .
 \label{eq:msignal}
\end{align}

The decay $\psip\to Y_n\bar{Y}_n e^+e^-$, where $Y_n$ denotes the neutral isospin partner of the charged hyperon $Y$, is dominated by the type-X diagrams since $Y_n$ and $\bar{Y}_n$ are charge neutral.\footnote{{Due to charge neutrality, the leading coupling of $Y_n$ to the photon vanishes. The EMFF of the neutral hyperon starts from the mean square charge radius term, and is thus much smaller than that of the charged hyperon in the small $Q^2$ region. When precise data become available, such contributions can, in principle, be included in the analysis.}}
Because the two-gluon exchange is isospin symmetric, the type-X contribution to the decay rate of $\psip\to Y\bar{Y} e^+e^-$ should equal to that of $\psip\to Y_n\bar{Y}_n e^+e^-$ up to tiny corrections. Therefore, the contribution from $\left\lvert {\cal M}_X \right\rvert^2$ in Eq.~\eqref{eq:amplitude} to the differential decay rate of $\psip\to Y\bar{Y} e^+e^-$ can be eliminated by subtracting the differential decay rate of $\psip\to Y_n\bar{Y}_n e^+e^-$ from that of $\psip\to Y\bar{Y} e^+e^-$ (since both charged and neutral $\Sigma$ and $\Xi$ hyperons can be reconstructed from proton, $\pi^-$ and photon(s), they can be easily detected).
Then the hyperon EMFFs are contained in all the left terms, denoted as $\left\lvert {\cal M}_{\rm signal} \right\rvert^2$, in the decay rate after subtraction.

However, not only the hyperon EMFFs but also the transition form factors from excited resonances to $\gamma^*Y/\bar Y$ can contribute to diagrams (c) and (d). Therefore, it is crucial to identify a kinematic region where the ground state octet hyperon contribution dominates over such background.
When the $Y\bar Y$ invariant mass $m_{Y\bar Y}$ is large, the energy-momentum conservation forces both $m_{Y\gamma^*}$ and $m_{\bar Y \gamma^*}$ to be small, and thus the hyperon/anti-hyperon pole should dominate the type-A and type-B amplitudes.
In the subsequent analysis, we will show that this is indeed the case. Given that the $\Sigma(1385)^+$ and $\Xi(1530)^-$ are the lowest-lying resonance that can couple to $\Sigma^+\gamma^*$ and $\Xi^-\gamma^*$, respectively, we investigate the $\Sigma(1385)^+$ and $\Xi(1530)^-$ contributions to pinpoint the kinematic regions dominated by the $\Sigma^+$ and $\Xi^-$ poles, respectively. We define the following differential ratio
\begin{equation}\label{eq:raND}
	\frac{d R}{d m_{e^+e^-} d m_{Y\bar{Y}}}=\frac{\int { d\cos\theta^*d\cos\theta^\prime d\phi}\, d\Gamma_{A+B}^Y}{\int  {d\cos\theta^*d\cos\theta^\prime d\phi}\,d\Gamma^{Y+Y^*}_{A+B}}~.
\end{equation}

The amplitude of 
$\psip(p_0) \to e^-(p_1)+e^+(p_2)+Y(p_3)+\bar{Y}(p_4)$
can be written as,
\begin{align}
	i{\cal M}_{(i)}&=H^\mu_{(i)} L^\nu \frac{-i g_{\mu\nu}}{q^2}, 
\end{align}
where $q=p_0-p_3-p_4$ the four-momentum of the virtual photon, $(i) = $(a), (b), (c) or (d) refers to the corresponding diagram in Fig.~\ref{fig:diagram},
and the whole decay amplitude is decomposed into a hadronic ($H^\mu$) and a leptonic ($L^\nu$) contribution. 
The leptonic operator is given by
\begin{equation}
	L^\nu=-i e \bar{u}_{s_{e^-}}(p_1)\gamma^\nu v_{s_{e^+}}(p_2)~.
\end{equation}
The EMFFs are contained in the hadronic operators, for which the vertices $\psip Y\bar{Y}$, $\psip Y^*\bar{Y}$ and $\psip Y\bar{Y}^*$ are constructed within the orbital-spin partial-wave framework, incorporating both $S$- and $D$-wave contributions~\cite{Zou:2002ar,Wu:2021yfv,Jing:2023rwz}. The $\gamma Y^* Y$ interaction is parameterized with a magnetic dipole form factor $g_M^{Y^*}$ normalized to the partial width of $Y^* \to Y +\gamma$~\cite{Pascalutsa:2006up}, and the electric quadrupole and Coulomb quadrupole terms have been neglected as they were found to be at the percent level for the analogous $\Delta\to N\gamma^*$~\cite{Tiator:2011pw,Guttmann:2012sq}. Technical details, including gauge invariance and off-shell effects~\cite{Gross:1987bu,Scherer:1996ux,Haberzettl:2006bn}, are provided in the Supplemental Material~\cite{SM}. 

\begin{figure}[t]
	\centering
	\includegraphics*[width=0.49\linewidth,angle=0]{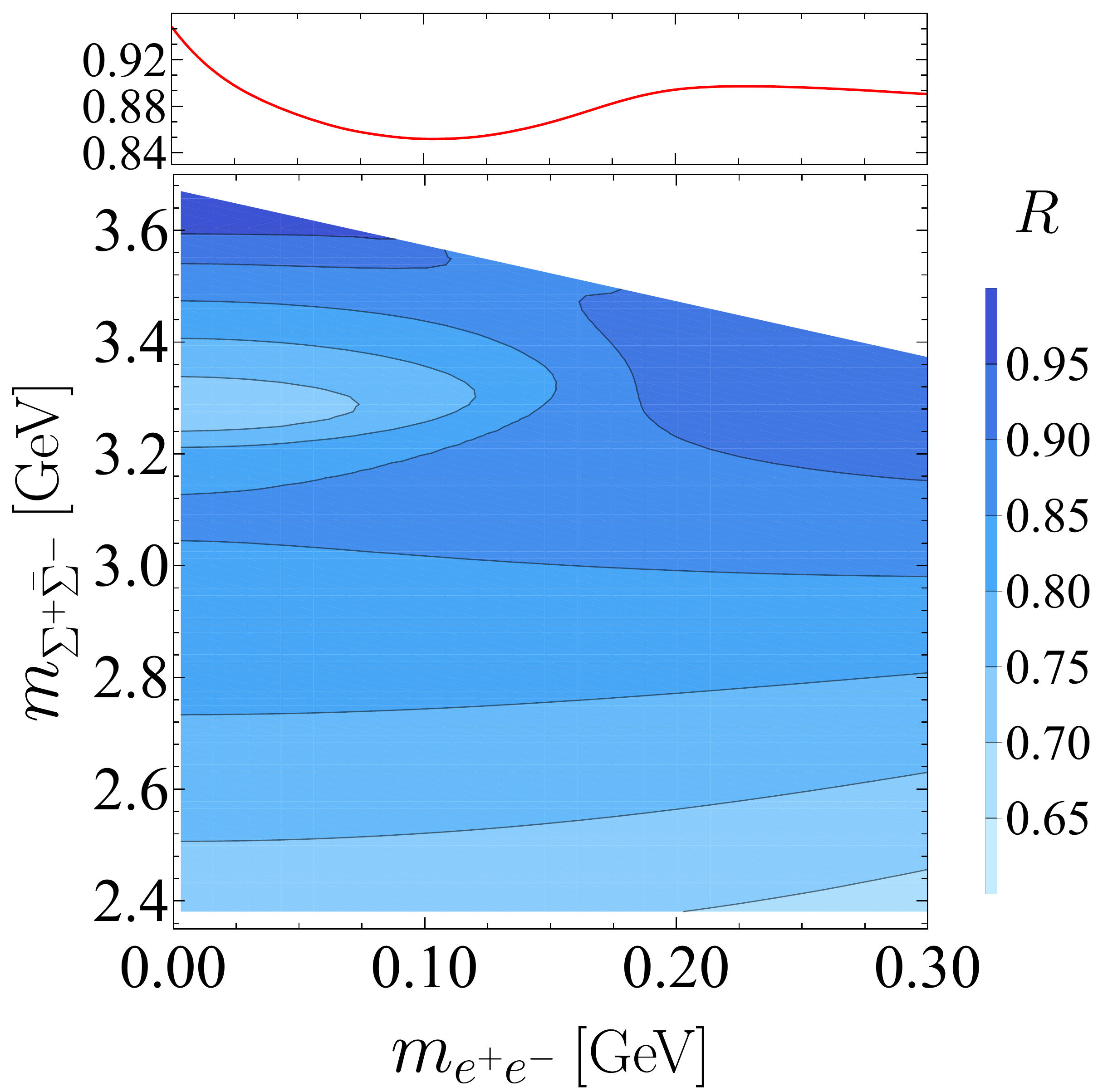}\hfill 
	\includegraphics*[width=0.485\linewidth,angle=0]{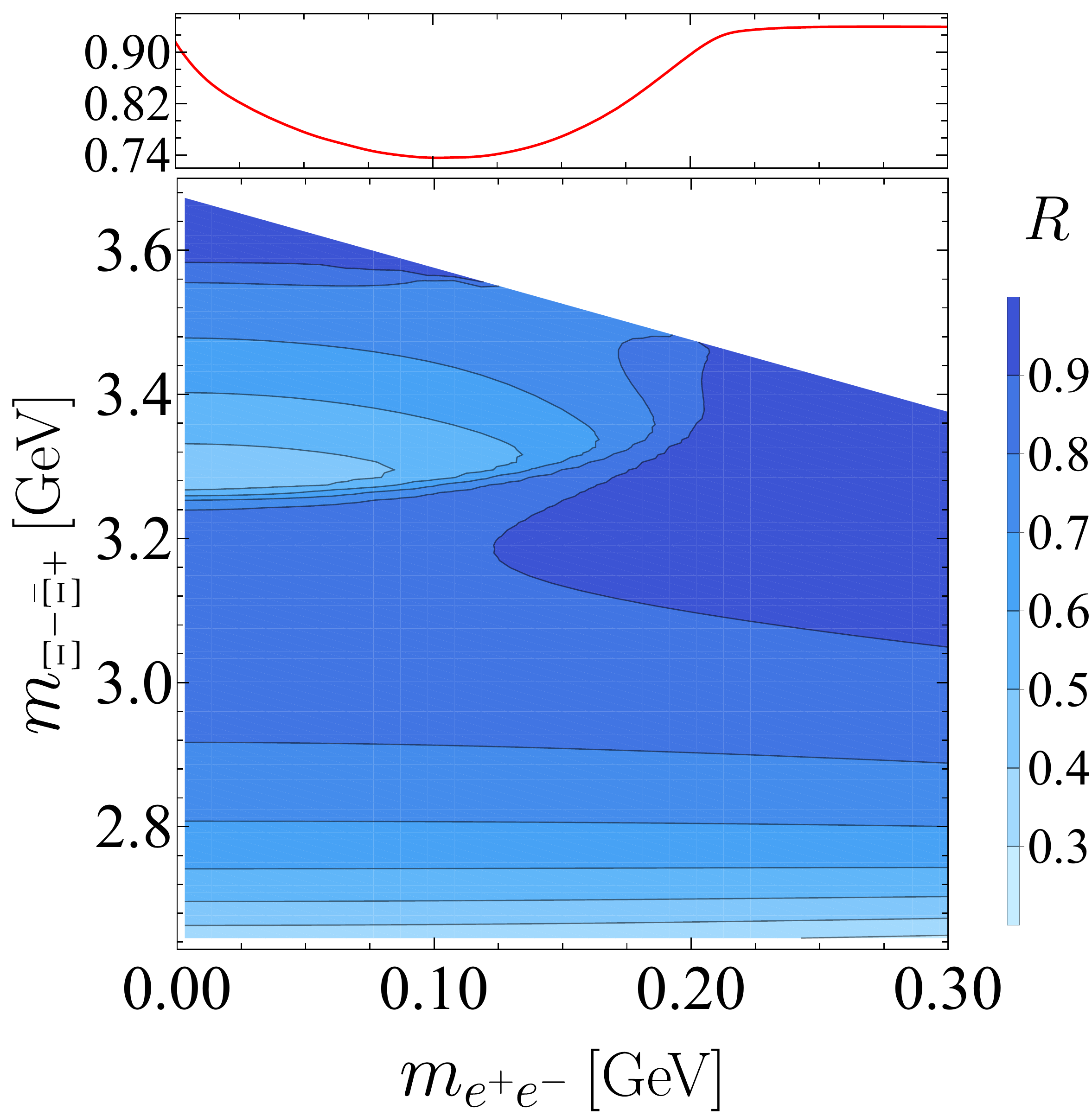}
	\caption{
	Differential ratio $R$ defined in Eq.~\eqref{eq:raND} for $\Sigma^+$ (left) and $\Xi^-$.
  The projected subplots show the corresponding ratio with $m_{\Sigma^+\bar \Sigma^-}$ and $m_{\Xi^{-} \bar{\Xi}^{+}}$ integrated out from 2.8 and 2.9~GeV, respectively.
	}
	\label{fig:NDratio}
\end{figure} 
Using the dipole hypothesis for the elastic hyperon EMFFs and setting the charge radius to $0.8\ {\rm fm}$, in the ballpark of the proton and $\Sigma^-$ radii, we present the resulting ratio $R$, as defined in Eq.~\eqref{eq:raND}, in Fig.~\ref{fig:NDratio} for $\Sigma^+$ and $\Xi^-$ hyperons. 
It is shown that 
for $m_{\Sigma^+\bar{\Sigma}^-}>2.8\,{\rm GeV}$ ($m_{\Xi^-\bar{\Xi}^+}>2.9\,{\rm GeV}$), the $\Sigma^+$ ($\Xi^-$) pole contributes at least 85\% (74\%) to the type-A and type-B mechanisms,
which is similar with the proton case where there are no any visible $m_{p\gamma}$ or $m_{\bar p\gamma }$ bands in the Dalitz plot of $\jpsi\to p\bar{p}\gamma $~\cite{Kappert:2022fox}.
We checked that this ratio is insensitive to the input hyperon charge radius. 

Therefore, it is feasible to extract the $\Sigma^+$ and $\Xi^-$ EMFFs from the kinematic region of $m_{\Sigma^+\bar{\Sigma}^-}>2.8\,{\rm GeV}$ of the decay $\psip \to \Sigma^+\bar{\Sigma}^-e^+e^-$ and $m_{\Xi^-\bar{\Xi}^+}>2.9\,{\rm GeV}$ of the decay $\psip \to \Xi^-\bar{\Xi}^+e^+e^-$, respectively. The process can be measured at \bes{} and at the potential STCF with two-orders-of-magnitude more events. One notable advantage is the ability to reach an unprecedented low momentum transfer squared ($4m_{e}^2$) through this decay, which offers a way for measuring of the unknown charge radii of the $\Sigma^+$ and $\Xi^-$.

\section{Sensitivity to the charge radii of $\Sigma^+$ and $\Xi^-$}

By subtracting the type-X contribution using the $\Sigma^0\bar{\Sigma}^0$ and $\Xi^0\bar{\Xi}^0$ measurements, as discussed above, and employing a parameterization for the elastic EMFFs of $\Sigma^+$ and $\Xi^-$, the differential decay rates, obtained within the previously described formalism, can be fitted to the event distribution in the low $m_{e^+e^-}^2$ region to extract the shape parameters of $\Sigma^+$ and $\Xi^-$ hyperons, respectively.
It is instructive to investigate the sensitivity of the $e^+e^-$ invariant-mass distribution to the hyperon charge radius. We now calculate the $m_{e^+e^-}$ distribution in the range of $m^2_{e^+e^-}<0.1\,{\rm GeV^2}$ with $m_{\Sigma^+\bar{\Sigma}^-}>2.8\,{\rm GeV}$ for the decay $\psip \to \Sigma^+\bar{\Sigma}^-e^+e^-$ ($m_{\Xi^-\bar{\Xi}^+}>2.9\,{\rm GeV}$ for the decay $\psip \to \Xi^-\bar{\Xi}^+e^+e^-$) using a dipole EMFF as input. Specifically, we utilize the simple dipole form for $G_E$ in conjunction with the Kelly description~\cite{Kelly:2004hm} for $G_M$. 

In a full experimental analysis, both the type-X contribution can be accounted for by a partial wave analysis similar to that performed in Ref.~\cite{Kappert:2022fox}. 
Here, to be concrete, we construct the type-X amplitude to get the signal amplitude of Eq.~\eqref{eq:msignal}. Prior experimental studies indicate that the type-X contribution in the suggested kinematic region is saturated by four charmonia: $\eta_c$, $\chi_{c0}$, $\chi_{c1}$ and $\chi_{c2}$~\cite{CLEO:2010fre}. 
Transitions among the initial $\psip$ and these $X$ charmonia are consistently parameterized with two couplings: one for $\eta_c$ and the other for the three $\chi_{cJ}$ states, which is facilitated by an effective Lagrangian grounded in heavy quark spin symmetry~\cite{DeFazio:2008xq,Wang:2011rt}. Moreover, a phenomenological transition form factor $f_{\psi}(q^2)=1/(1-q^2/\Lambda_X^2)$ with $\Lambda_X=m_{\omega}$ is introduced, which has been widely used in analyzing the $\jpsi$ Dalitz decay~\cite{Fu:2011yy,BESIII:2021xoh,BESIII:2022jde}. Subsequent decays of $X\to Y\bar Y$ are described within the orbital-spin partial-wave framework~\cite{Zou:2002ar,Jing:2023rwz}, similar to that of $\psip\to Y\bar Y$ vertices. All the effective couplings are determined by the relevant partial widths listed in the Review of Particle Physics (RPP)~\cite{ParticleDataGroup:2022pth}. Technical details can be found in the Supplemental Material~\cite{SM}. 

\begin{figure}[tb]
	\centering
	\includegraphics*[width=0.49\linewidth,angle=0]{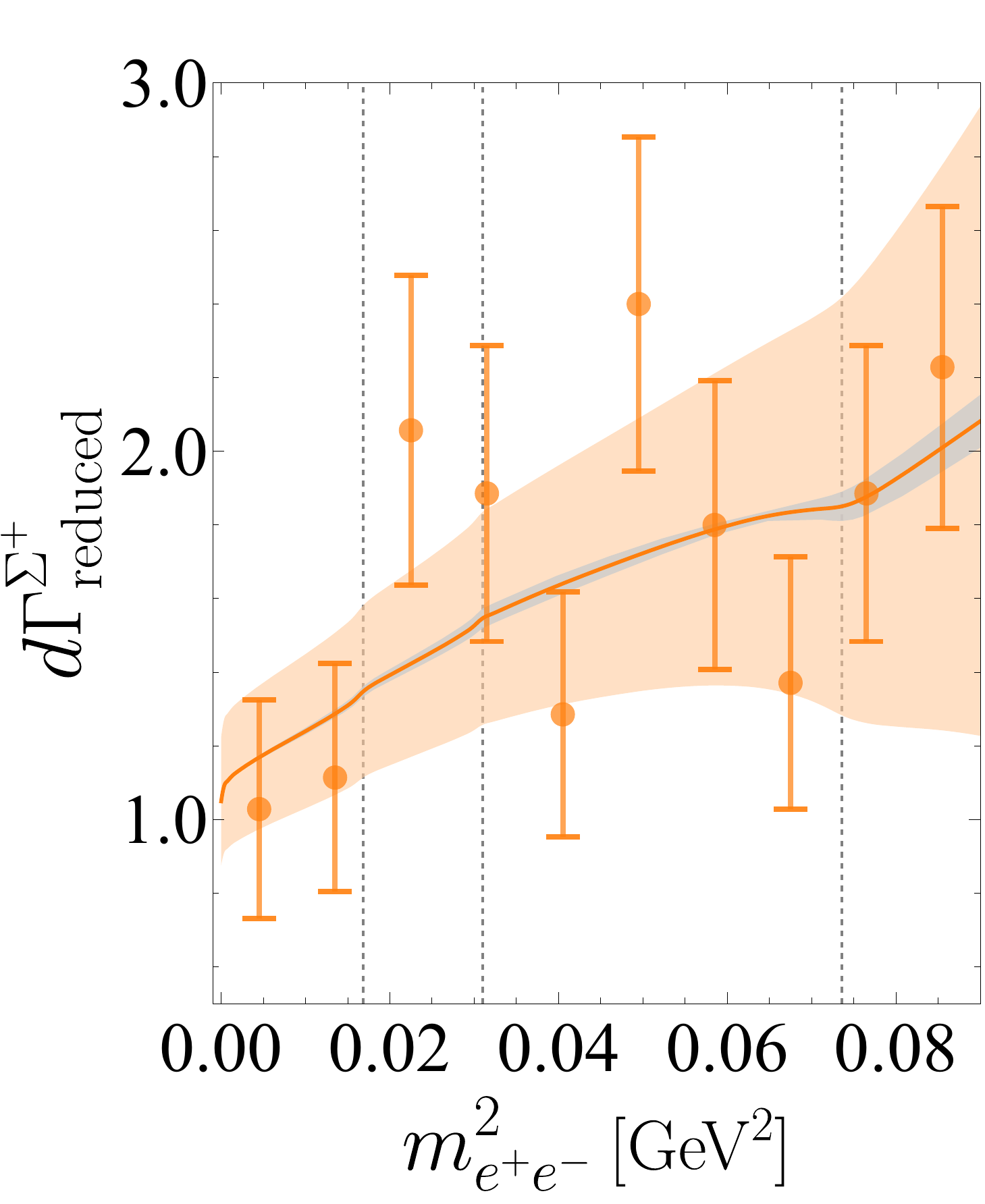}\
	\includegraphics*[width=0.49\linewidth,angle=0]{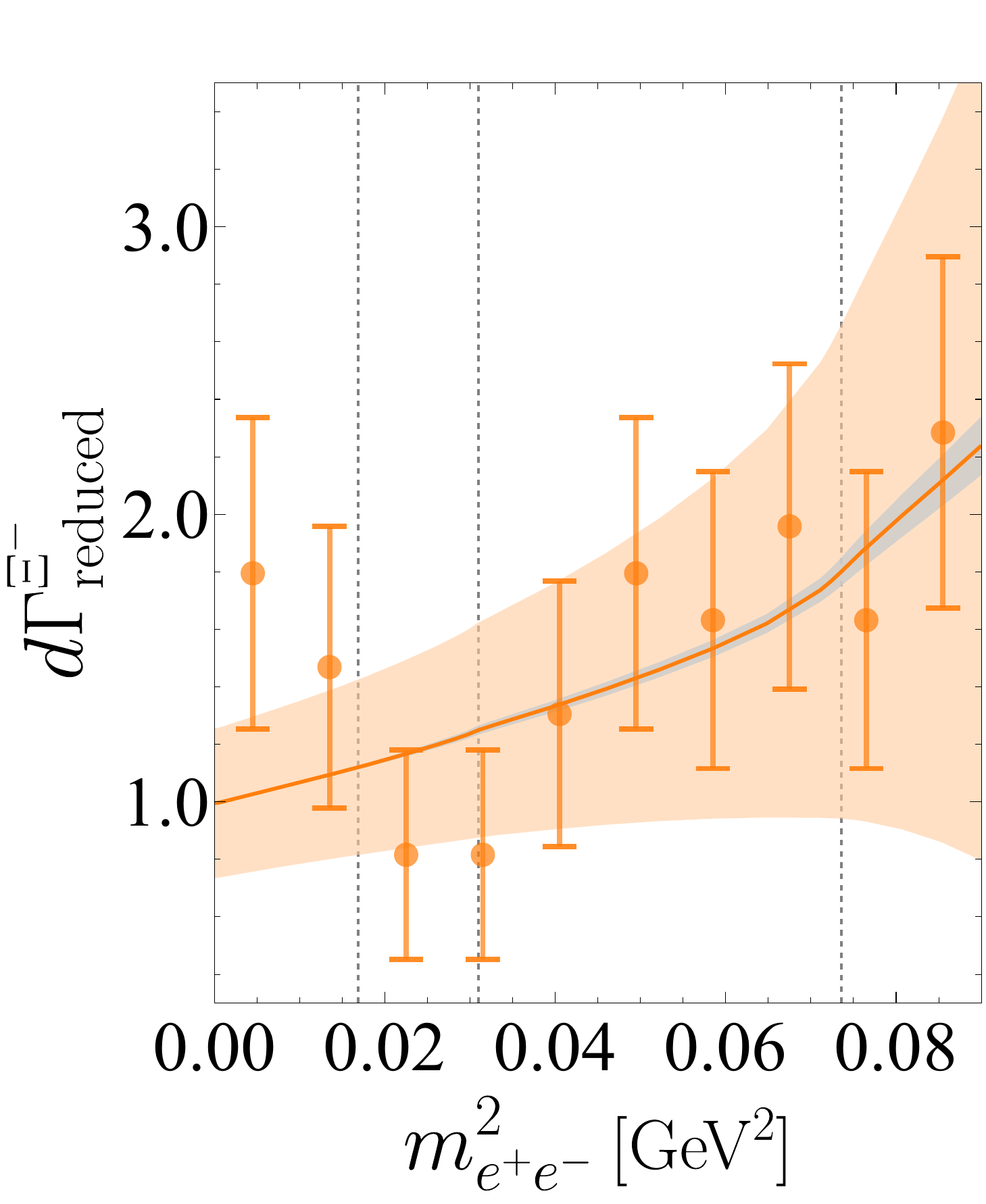}
	\caption{Fit to $200$ synthetic Monte Carlo events using 0.8~fm as the input radius value: $\Sigma^+$ (left) and $\Xi^-$ (right). The extracted radii from fitting to the two data sets are $1.02(23)$~fm for $\Sigma^+$ and $1.13(23)$~fm for $\Xi^-$, respectively. Three vertical dashed lines denote the right endpoints of the three $\chi_{cJ}$ resonances (from left to right: $\chi_{c2}$, $\chi_{c1}$ and $\chi_{c0}$) projected onto the $m_{e^+e^-}$ axis. The gray bands show the effects of model parameters to the radii extractions, as detailed in the text.
	}
	\label{fig:mc-bes}
\end{figure}

\begin{figure}[tb]
	\centering
	\includegraphics*[width=0.49\linewidth,angle=0]{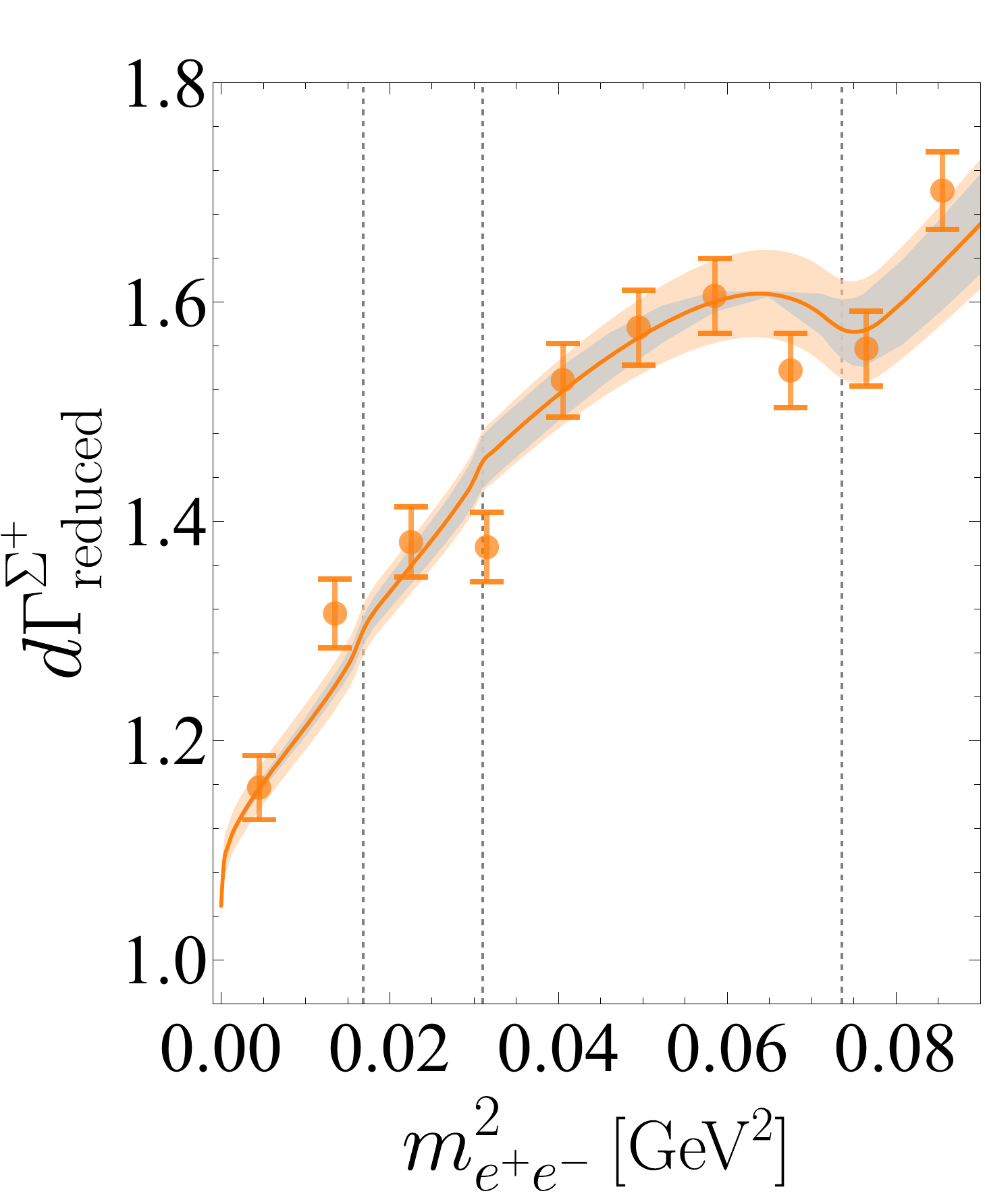}\
	\includegraphics*[width=0.49\linewidth,angle=0]{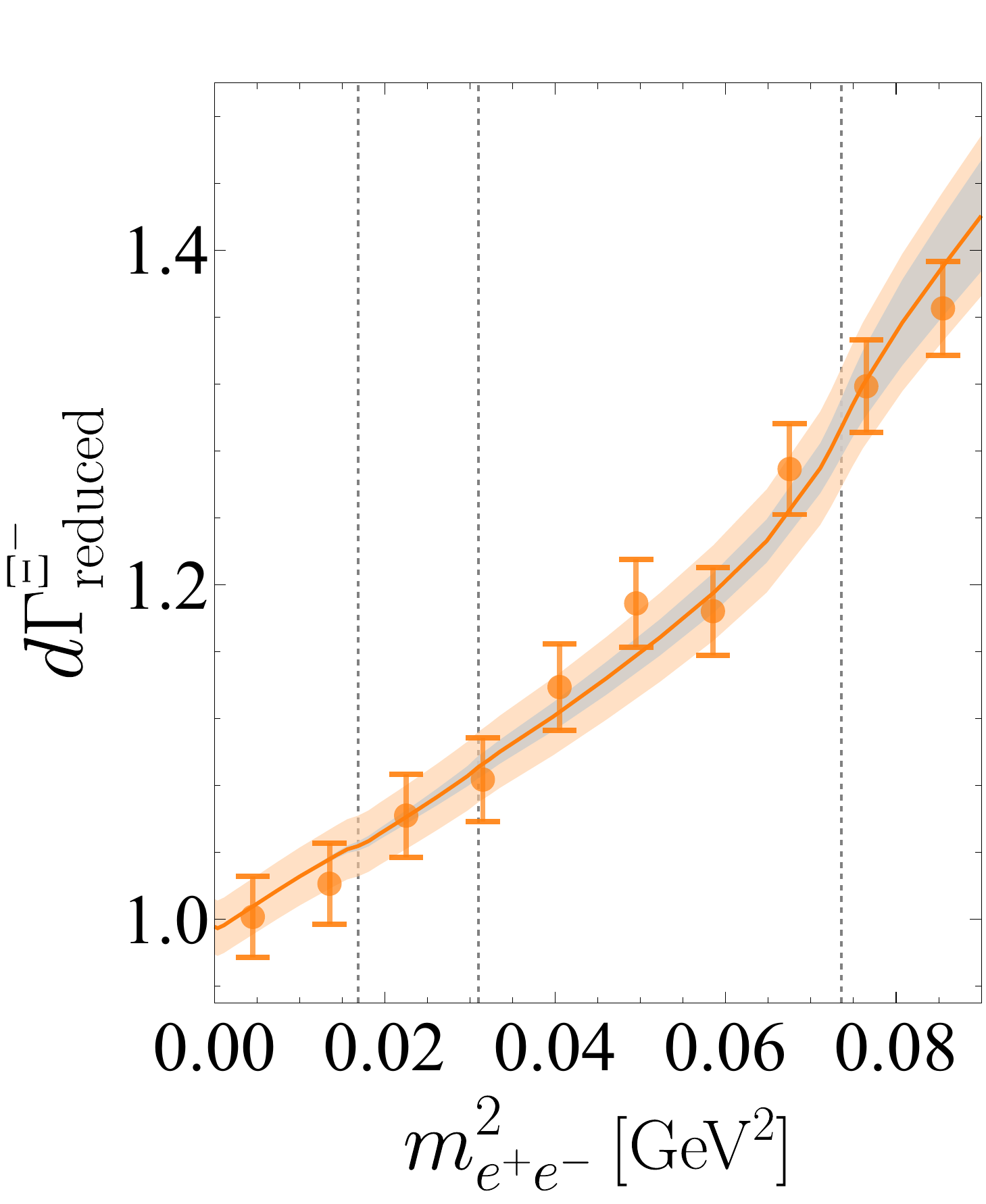}
	\caption{Fit to $2\times10^4$ synthetic Monte Carlo events using 0.8~fm as the input radius value: $\Sigma^+$ (left) and $\Xi^-$ (right). The extracted radii from fitting to the two data sets are $0.78(4)$~fm for $\Sigma^+$ and $0.80(3)$~fm for $\Xi^-$, respectively. For notations, refer to Fig.~\ref{fig:mc-bes}.
	}
	\label{fig:mc-stcf}
\end{figure}

First, we estimate the expected event yields for the radii extractions of the $\Sigma^+$ and $\Xi^-$ from the $\psip$ data sets at \bes{} and STCF by integrating the differential decay rate given in Eq.~\eqref{eq:dcs} over the relevant kinematic region of interest. It is found that there would be around $\mathcal{O}(10^2)$ events for both the $\Sigma^+$ and $\Xi^-$ cases at \bes{}, and about 300 times more for both cases at STCF. To better investigate the sensitivity to charge radii of $\Sigma^+$ and $\Xi^-$, we generate synthetic data following the distribution of the dipole form factor (for simplicity) for the $\Sigma^+$ and $\Xi^-$ with $r_E^{\Sigma^+}=r_E^{\Xi^-}=0.80~{\rm fm}$ using the von Neumann rejection method.
We generate two sets of samples with $2\times10^2$ and $2\times10^4$ events, corresponding to simulations of the BES\Rom{3}~\cite{BESIII:2020nme} and STCF~\cite{Achasov:2023gey} experiments. 
Fitting to the synthetic data leads to $1.02(23)$~fm for $\Sigma^+$ and $1.13(23)$~fm for $\Xi^-$ in the context of the \bes{} experiment, see Fig.~\ref{fig:mc-bes}. Results for the STCF experiment are presented in Fig.~\ref{fig:mc-stcf}, yielding $0.78(4)$~fm for $\Sigma^+$ and $0.80(3)$~fm for $\Xi^-$. Notice that the point-like ansatz for $\Sigma^+$ and $\Xi^-$ is used as a reference in Fig.~\ref{fig:mc-bes} and Fig.~\ref{fig:mc-stcf}. Consequently, effects from sources other than the elastic EMFFs of hyperons get largely neutralized. The reduced invariant mass distribution of $e^+e^-$ exhibits a kink behavior at the right endpoints of the three $\chi_{cJ}$ resonances when projected onto the $m_{e^+e^-}$ axis, attributed to their narrow widths.

We also test the robustness of the simulation results by conservatively varying the involved parameter values, which can be fixed in a real analysis of the experimental data. 
The gray bands in Fig.~\ref{fig:mc-bes} and Fig.~\ref{fig:mc-stcf} count the effects of amplitude parameters to the radii extractions, which are obtained by varying the $S/D$ ratio of the $\psip Y\bar{Y}$ vertex by 20\% (the uncertainty as determined in Ref.~\cite{Wu:2021yfv} is only 2\%), varying the charmonium transition couplings by 15\% and $\Lambda_X$ in the transition form factors $f_\psi(q^2)$ by 0.2~GeV. That the gray bands are within the $1\sigma$ region of the optimal charge radii suggests that the charge radius of $\Sigma^+$ or $\Xi^-$ predominates over these amplitude parameters in dictating the $e^+e^-$ invariant-mass distribution in the low $m_{e^+e^-}$ region, given the current experimental precision. As event counts increase, a combined fit incorporating all amplitude parameters and the charge radius becomes essential for a precise radius extraction using the proposed methodology.

This novel approach may also be checked by extracting the proton charge radius from the analogous Dalitz decay $\jpsi \to p\bar{p}e^+e^-$ at the \bes{} and STCF experiments; details of the applicability study can be found in the Supplemental Material~\cite{SM}.

\section{Summary and prospect}

In this Letter, we propose to measure the charge radii of charge hyperons $\Sigma^+$ and $\Xi^-$ from the time-like region of the Dalitz decays $\psip \to Y\bar Y e^+e^-$. The $-Q^2$ value can reach $\sim 4m_e^2=1.05\times 10^{-6}$~GeV$^2$, which is unprecedentedly low even for the proton charge radius measurements, making use of the advantage that the electron and positron with transverse momenta can be detected with very efficiency.

The optimal kinematic region for extracting the elastic EMFFs of $\Sigma^+$ and $\Xi^-$ is identified to be $m_{\Sigma^+\bar{\Sigma}^-}>2.8\,{\rm GeV}$ for the decay $\psip \to \Sigma^+\bar{\Sigma}^-e^+e^-$ and $m_{\Xi^-\bar{\Xi}^+}>2.9\,{\rm GeV}$ for the decay $\psip \to \Xi^-\bar{\Xi}^+e^+e^-$, where the involved photon-baryon coupling is dominated by the elastic EMFFs of $\Sigma^+$ and $\Xi^-$ and a clean background subtraction can in principle be performed. The latter is achievable through measuring the decays into neutral hyperons $\psip \to \Sigma^0\bar{\Sigma}^0e^+e^-$ and $\psip \to \Xi^0\bar{\Xi}^0e^+e^-$ and subtracting their event distributions from those of $\psip \to \Sigma^+\bar{\Sigma}^-e^+e^-$ and $\psip \to \Xi^-\bar{\Xi}^+e^+e^-$, respectively.

From a Monte Carlo simulation, we find that already with the BESIII data sample of $\psi(2S)$, the charge radii of the $\Sigma^+$ and $\Xi^-$ can be extracted with an uncertainty of 20\% level.
The precision can be significantly improved with two orders of magnitude more events at STCF. 
{First measurements of the charge radii of $\Sigma^+$ and $\Xi^-$, which have not been achieved so far, are foreseen employing this novel approach.}

\bigskip

\acknowledgments 
We are grateful to Hai-Bo Li, Jian-Ping Ma, Wei-Zhi Xiong, Hai-Qing Zhou and Xiao-Rong Zhou for useful discussions.
YHL and UGM are grateful to the hospitality of the Institute of Theoretical Physics, Chinese Academy of Sciences (CAS), where part of the work was done.
This work is supported in part by the CAS under Grants No.~YSBR-101 and No.~XDB34030000; by the National Natural Science Foundation of China (NSFC) and the Deutsche Forschungsgemeinschaft (DFG) through the funds provided to the Sino-German Collaborative Research Center TRR110 “Symmetries and the Emergence of Structure in QCD” (NSFC Grant No. 12070131001, DFG Project-ID 196253076); by the NSFC under Grants No. 12125507, No. 11835015, and No. 12047503; by CAS through the President’s International
Fellowship Initiative (PIFI) under Grant No. 2025PD0022; and by the VolkswagenStiftung under Grant No. 93562.

\bibliography{refs}

\pagebreak

\,

\pagebreak

\begin{appendix}
\begin{onecolumngrid}

\section{Supplemental Material}
\subsection{A. Hadronic Operators}

Here, we give some details on the hadronic operators in Fig.~\ref{fig:diagram}.
The hadronic contributions in diagrams (c) and (d) from the octet baryon poles are given by
\begin{align}
	H_{(c)}^{Y,\mu}=&\,\bar{u}_{s_p}(p_3)\phnn^\mu(-q)S^{1/2}_+(p_0-p_4) \cnn^\nu(p_0-2p_4,p_0)\epsilon_{s_{\psip}, \nu}(p_0) v_{s_{\bar{p}}}(p_4), \notag \\
	H_{(d)}^{Y,\mu}=&-\bar{u}_{s_p}(p_3)\cnn^\nu(2p_3-p_0,p_0)\epsilon_{s_{\psip}, \nu}(p_0) {S}^{1/2}_-(p_0-p_3)\phnn^\mu(-q) v_{s_{\bar{p}}}(p_4).
\end{align} 
For the contributions from the lowest excited decuplet resonances,
the corresponding hadronic operators take the form 
\begin{align}
	H_{(c)}^{\Delta,\mu}=&\,\bar{u}_{s_p}(p_3)\phnd^{\alpha\mu}(-q,p_0-p_4)S^{3/2}_{\alpha\beta,+}(p_0-p_4) \cnd^{\beta\nu}(p_0-2p_4,p_0)\epsilon_{s_{\psip}, \nu}(p_0) v_{s_{\bar{p}}}(p_4), \notag\\
	H_{(d)}^{\Delta,\mu}=&\,\bar{u}_{s_p}(p_3)\cnd^{\alpha\nu}(2p_3-p_0,p_0)\epsilon_{s_{\psip}, \nu}(p_0) {S}^{3/2}_{\alpha\beta,-}(p_0-p_3)\phnd^{\beta\mu}(-q,p_0-p_3) v_{s_{\bar{p}}}(p_4).
\end{align}
Here, $s_i$ denotes the polarization component of particle $i$, and $S^{1/2}_\pm$ and $S^{3/2}_\pm$ represent the $Y/\bar Y$ and $Y^*/\bar{Y}^*$ propagators, respectively ($+$ and $-$ are for the baryon and anti-baryon, respectively), 
\begin{align}
	S^{1/2}_\pm(p)=&\,\frac{i(\slashed{p}\pm m_Y)}{p^2-m_Y^2+i \epsilon},\notag \\
	S_{\mu\nu, \pm}^{3/2}(p)=&\,\frac{-i(\slashed{p}\pm m_{Y^*})}{p^2-m_\Delta^2+i m_{Y^*}\Gamma_{Y^*}}\bigg(g_{\mu\nu}-\frac13\gamma_\mu\gamma_\nu \pm\frac1{3m_{Y^*}}(p_\mu\gamma_\nu-p_\nu\gamma_\mu)-\frac2{3m_{Y^*}^2}p_\mu p_\nu\bigg).
\end{align}
In addition, $\Gamma_{\gamma Y Y}$ and $\Gamma_{\gamma Y^* Y}$ denote the electromagnetic vertices in the type-A and type-B diagrams in Fig.~\ref{fig:diagram}, which can be parameterized in terms of three EMFFs, that is, the Dirac ($F_1$) and Pauli ($F_2$) form factors of the octet hyperons and the magnetic dipole form factor ($g_M^{Y^*}$) characterizing the strength of the magnetic dipole (M1) $Y^*\to Y\gamma^*$ transition~\cite{Pascalutsa:2006up}, as
\begin{align}
	&\Gamma_{\gamma Y Y}^\mu(q)=i e \left(\gamma^\mu F_1(q^2)+\frac{i\sigma^{\mu\nu}}{2m_Y}q_\nu F_2(q^2)\right),\label{eq:enNN}\\
	&\Gamma_{\gamma Y^* Y}^{\alpha\mu}(q,p_{Y^*})=ie\sqrt{\frac23}\frac{3(m_Y+m_{Y^*})}{2m_Y((m_{Y^*}+m_Y)^2-q^2)}g_M^{Y^*}(q^2)\epsilon^{\alpha\mu\rho\sigma}p_{{Y^*},\rho}q_{\sigma}.\label{eq:emND}
\end{align}
Here, the Lorentz index $\alpha$ contracts with the vector-spinor of the ${Y^*}$ decuplet baryon while $\mu$ contracts with the photon polarization vector. The electric quadrupole and Coulomb quadrupole terms for $Y^*\to Y\gamma^*$ have been neglected in Eq.~\eqref{eq:emND} as done for the amplitude $\Delta\to N\gamma^*$ in Ref.~\cite{Guttmann:2012sq} since they contribute only at the percent level~\cite{Tiator:2011pw}.

The vertices $\Gamma_{\psip Y\bar{Y}}$ and $\Gamma_{\psip {Y^*}\bar{Y}}$ are constructed in the Lorentz covariant orbital-spin scheme~\cite{Zou:2002ar,Jing:2023rwz} as
\begin{align}
	\Gamma_{\psip Y\bar{Y}}^\mu(r,p_0)=&\,g_S \left(\gamma^\mu-\frac{r^\mu}{M_{\psip}+2m_Y}\right)+g_D e^{i\delta_1}\left(\gamma_\nu-\frac{r_\nu}{M_{\psip}+2m_Y}\right)t^{\mu\nu},\label{eq:JpsiN}\\
	\Gamma_{\psip {Y^*}\bar{Y}}^{\mu\alpha}(r,p_0)=&\,f_S \gamma_5 g^{\mu\alpha} +f_D e^{i\delta_2}\gamma_5 t^{\mu\alpha},\label{eq:JpsiND}
\end{align}
where $r$ is the relative four-momentum between the baryon and anti-baryon of the $\psip Y\bar{Y}$ vertex, and $t^{\mu\nu}$ denotes the $D$-wave tensor amplitude in the covariant orbital-spin scheme, given by
\begin{align}
	t^{\mu\nu}(r,p_0)=&\,r^\mu r^\nu-\frac{r^2}{3}g^{\mu\nu}+\frac{r^2}{3 M_{\psip}^2}p_0^\mu p_0^\nu+\frac{(p_0\cdot r)^2}{3 M_{\psip}^2}\left(g^{\mu\nu}+\frac2{3M_{\psip}^2}p_0^\mu p_0^\nu\right)-\frac{p_0\cdot r}{ M_{\psip}^2}\left(r^\mu p_0^\nu -r^\nu p_0^\mu\right).
\end{align}
All the couplings $g_S$, $g_D$, $f_S$ and $f_D$ are real-valued with the relative phases between the $S$- and $D$-wave components manifested by two phase factors $\delta_1$ and $\delta_2$, respectively. {Note that, however, gauge invariance is broken when one combines the $\psip Y\bar{Y}$ effective vertex of Eq.~\eqref{eq:JpsiN} with the $\gamma YY$ amplitude in Eq.~\eqref{eq:enNN}. A contact term for the four-point $\psip \gamma Y\bar{Y}$ amplitude has to be introduced to restore the gauge invariance for the given process, see next subsection for details.}

As introduced in the main text, the type-X contribution to $\psip \to Y\bar{Y}e^+e^-$ within the targeted kinematic region is dominated by four charmonia: $\eta_c$, $\chi_{c0}$, $\chi_{c1}$ and $\chi_{c2}$, depicted solely by diagram (a) in Fig.~\ref{fig:diagram}. 
To parameterize the amplitudes of these charmonium states, we utilize an effective Lagrangian that incorporates heavy quark spin symmetry for $\psip \gamma X$ vertices~\cite{DeFazio:2008xq,Wang:2011rt}. This is combined with the description of the subsequent $X\to Y\bar Y$ decays within the orbital-spin partial-wave framework. Then the type-X hadronic operators read
\begin{align}
	H_{(a)}^{\mu}=&\,f_\psi(q_X^2)\bigg\{\sum_{X_i=\eta_c,\chi_{c0}}\,\bar{u}_{s_p}(p_3)\Gamma_{X_i Y\bar{Y}}(p_3-p_4, p_X)v_{s_{\bar{p}}}(p_4)G_0(p_X) \Gamma_{\psip X_i \gamma}^{\mu\nu}(q_X)\epsilon_{s_{\psip}, \nu}(p_0)\ \notag\\
	&+\bar{u}_{s_p}(p_3)\Gamma_{\chi_{c1} Y\bar{Y}}^\rho(p_3-p_4,p_X)v_{s_{\bar{p}}}(p_4)G_{1, \rho\alpha}(p_X) \Gamma_{\psip\chi_{c1} \gamma}^{\alpha\beta\mu}(q_X)\epsilon_{s_{\psip}, \beta}(p_0)\ \notag\\
	&+\bar{u}_{s_p}(p_3)\Gamma_{\chi_{c2} Y\bar{Y}}^{\rho\sigma}(p_3-p_4,p_X)v_{s_{\bar{p}}}(p_4)G_{2, \rho\sigma\alpha\beta}(p_X) \Gamma_{\psip \chi_{c2} \gamma}^{\alpha\beta\nu\mu}( q_X)\epsilon_{s_{\psip}, \nu}(p_0)\bigg\},\label{eq:hadronX}
\end{align}
with $p_X = p_3+p_4$, $q_X=p_0-p_X$, and the $\psip \gamma X$ vertices parameterized by two couplings $g_X$ and $f_X$,
\begin{align}
	\Gamma_{\psip \eta_c \gamma}^{\mu\nu}(q)&=g_X\epsilon^{\mu\nu\sigma \rho}q_{X,\rho}v_{\sigma} ,\notag\\
	\Gamma_{\psip \chi_{c0} \gamma}^{\mu\nu}( q)&=2\sqrt{3}f_X\left(v^\nu q^\mu-g^{\mu\nu} q\cdot v\right) ,\notag\\
	\Gamma_{\psip \chi_{c1} \gamma}^{\mu\rho\sigma}( q)&=3\sqrt{2}f_X\left(\epsilon^{\rho\mu\alpha\nu}q_{\alpha}v_\nu v^\sigma-\epsilon^{\sigma\rho\mu\nu}v_{\nu}q\cdot v\right) ,\notag\\
	\Gamma_{\psip \chi_{c2} \gamma}^{\mu\nu\rho\sigma}( q)&=6f_X\left(q^\mu v^\sigma (g^{\nu\rho}-v^\rho v^\nu)-q\cdot v(g^{\sigma\mu}g^{\nu\rho}-g^{\sigma\mu}v^\nu v^\rho)\right). 
	\label{eq:psiXph}
\end{align}
Here, $v=p_0/M_{\psip}$ denotes the heavy quark velocity. The $XY\bar Y$ vertices are constructed within the orbital-spin partial-wave framework as
 \begin{align}
 	\Gamma_{\eta_c Y \bar Y}(r, p)&=g_{\eta_c}^Y \gamma_5 ,\notag\\
 	\Gamma_{ \chi_{c0} Y \bar Y}(r, p)&=g_{\chi_{c0}}^Y B_\mu(r) r_\nu \tilde{g}^{\mu\nu}(p)  ,\notag\\
 	\Gamma_{ \chi_{c1} Y \bar Y}^{\mu}(r, p)&=g_{\chi_{c1}}^Y \epsilon^{\mu\alpha\nu\beta}  B_\alpha(r) v^\prime_\beta \tilde{g}_{\nu\rho}(p) r^\rho,\notag\\
 	\Gamma_{ \chi_{c2} Y \bar Y}^{\mu\nu}(r, p)&=g_{\chi_{c2}}^Y B^\mu(r)\tilde{g}^{\nu\rho}(p) r_\rho,
 	\label{eq:XNN}
 \end{align}
where $B_\mu(r)=\gamma_\mu-r^\mu/(M_X+2m_Y)$ and $\tilde{g}^{\mu\nu}(p)=g^{\mu\nu}-p^\mu p^\nu/p^2$, $r$ is the relative four-momentum between the baryon and anti-baryon of the $X Y\bar{Y}$ vertex, and $v^\prime=p_{\chi_{c1}}/M_{\chi_{c1}}$. Note that $G$ denotes the propagators for these intermediate $X$ resonances, whose expressions are given by 
\begin{align}
	G_0(p)&=\frac{i}{p^2-M_X^2+i M_X \Gamma_X} , \quad G_1^{\mu\nu}(p)=\frac{i(-g^{\mu\nu}+p^\mu p^\nu /M_X^2)}{p^2-M_X^2+i M_X \Gamma_X}, \notag\\
	G_2^{\mu\nu\alpha\beta}(p)&=\frac{i}{p^2-M_X^2+i M_X \Gamma_X}\left(\frac12(V^{\mu\alpha}(p)V^{\nu\beta}(p)+V^{\mu\beta}(p)V^{\nu\alpha})-\frac13 V^{\mu\nu}(p)V^{\alpha\beta}(p)\right),
	\label{eq:Xppg}
\end{align}
with $V^{\mu\nu}(p)=-g^{\mu\nu}+p^\mu p^\nu/M_X^2$. Moreover, in the type-X hadronic operators given by Eq.~\eqref{eq:hadronX}, $f_\psi(q^2)=1/(1-q^2/\Lambda_X^2)$ is a phenomenological transition form factor that is widely used in analyzing the $\jpsi$ Dalitz decay~\cite{Fu:2011yy,BESIII:2021xoh,BESIII:2022jde}. We take $\Lambda_X=0.78\pm 0.20\ {\rm GeV}$ for the $\psip \to X\gamma^*$ transitions motivated by the vector meson dominance model. 

{The real-valued parameters contained in these hadronic operators are the $S$-wave and $D$-wave couplings $g_S$, $g_D$ and their relative phase $\delta_1$ within $\Gamma_{\psip Y\bar{Y}}$, and analogously, $f_S$, $f_D$ and $\delta_2$ in the vertex of $\Gamma_{\psip Y^*\bar{Y}}$. For the type-X contributions, we have two couplings $g_X$ and $f_X$ characterizing the radiative transitions between $\psip$ and $X$ charmonia, and additionally 9 effective couplings $g_X^Y$ contained in the $XY\bar Y$ vertices.} 
To determine both $S$-wave and $D$-wave couplings solely from the partial decay widths of $\psip \to Y\bar Y (Y^*\bar{Y})$, prior knowledge on the ratio $R_{D/S}\equiv g_D /g_S$ and the relative phase $\delta_i$ is necessary, which has been systematically studied in Ref.~\cite{Wu:2021yfv}. All remaining coupling constants can be straightforward inferred from the corresponding decay widths. Moreover, the elastic electromagnetic form factors of hyperons are normalized to their charge and magnetic moments given by RPP~\cite{ParticleDataGroup:2022pth}. The normalization of the transition form factor $g_M^{Y^*}$ is determined by the relevant radiative decay width of $Y^*\to Y+\gamma$. The numerical values for the involved coupling constants are presented in Table~\ref{Tab:c1}, Table~\ref{Tab:c2} and Table~\ref{Tab:c3}.
\begin{table}[tb]
	\centering
	\renewcommand\arraystretch{1.0}
	\caption{Coupling constants $f_X$, $g_X$ and $g_X^\prime$ for the $\psip \chi_{cJ}\gamma$, $\psip \eta_c\gamma$ and $\jpsi \eta_c\gamma$ vertices, respectively. Note that the 15\% uncertainty of $f_X$ comes from the three different $\chi_{cJ}$ values. And the same percentage uncertainty is also considered for $g_X$ and $g_X^\prime$. All these coupling constant are dimensionless. \label{Tab:c1}}
	\begin{ruledtabular}
		\begin{tabular}{cccccc}
			 $f_X(\chi_{c0})$ & $f_X(\chi_{c1})$ & $f_X(\chi_{c2})$	& $f_X$ & $g_X$ & $g_X^\prime$ \\
			\hline
			0.26  & 0.28  
			& 0.34   
			& $0.30\pm0.04$ & 0.045 & 0.65 
		\end{tabular}
	\end{ruledtabular}
\end{table}

\begin{table}[tb]
	\centering
	\renewcommand\arraystretch{1.0}
	\caption{Coupling constants $g_{\eta_c}^Y$ (dimensionless), $g_{\chi_{c0}}^Y$ (${\rm GeV^{-1}}$), $g_{\chi_{c1}}^Y$ (${\rm GeV^{-1}}$) and $g_{\chi_{c2}}^Y$ (${\rm GeV^{-1}}$) for the $XY\bar{Y}$ vertices. In addition, $g_{\eta_c}^p=0.022$ for the $\eta_c p\bar{p}$ vertex. \label{Tab:c2}}
	\begin{ruledtabular}
		\begin{tabular}{cccc}
			$g_{\eta_c}^\Sigma$ & $g_{\chi_{c0}}^\Sigma$ & $g_{\chi_{c1}}^\Sigma$ & $g_{\chi_{c2}}^\Sigma$ \\
			\hline
			0.031  & $2.9\times 10^{-3}$  
			& $2.6\times 10^{-4}$  & $5.2\times 10^{-4}$ \\
			\hline
			$g_{\eta_c}^\Xi$ & $g_{\chi_{c0}}^\Xi$ & $g_{\chi_{c1}}^\Xi$ & $g_{\chi_{c2}}^\Xi$ \\
			\hline
			0.023  & $3.5\times 10^{-3}$ 
			& $3.9\times 10^{-4}$  & $1.3\times 10^{-3}$ 
		\end{tabular}
	\end{ruledtabular}
\end{table}

\begin{table}[tb]
	\centering
	\renewcommand\arraystretch{1.0}
	\caption{Coupling constants $g_S$, $g_D$, $f_S$ and $f_D$ for a given  $\Gamma_S/\Gamma_{S+D}$ value. Note that a 20\% uncertainty in the ratio $R_{D/S}$ is considered to investigate the effects of model parameters on the radius extraction. \label{Tab:c3}}
	\begin{ruledtabular}
		\begin{tabular}{lcc c}
			$\Gamma_S/\Gamma_{S+D}$({88.3 \%})  
			& $\Sigma^+$ & $\Xi^-$ & $p$ \\
			\hline
			$g_S$ (dimensionless)  & $7.50\times10^{-4}$  
			& $8.54\times10^{-4}$  
			& $1.33\times10^{-3}$ \\
			$R_{D/S}\,({\rm GeV^{-2}})$	& 0.097 
			&0.117 &0.127	\\
			$\cos\delta_1$  & 0.944  
			& 0.944 & 0.862 \\
			$f_S$ (dimensionless)  	& $2.64\times10^{-4}$  
			& $1.70\times10^{-4}$ & $<3.64\times10^{-4}$	
		\end{tabular}
	\end{ruledtabular}
\end{table}

\subsection{B. Gauge Invariance}
The off-shell octet baryon momentum-dependent terms in the vertex $\Gamma_{\psip Y\bar{Y}}$ break  gauge invariance. In general, one can restore gauge invariance straightforwardly by replacing the momentum derivative acting on the octet hyperon field with the gauge-covariant derivative in the Lagrangian level, i.e., $\partial_\mu \to \partial- i e A_\mu $. Working at the amplitude level in practice, it leads to the following substitutions,
\begin{equation}
    k_Y^\mu\to k_Y^\mu+e A^\mu,\quad\quad k_{\bar{Y}}^\mu\to k_{\bar{Y}}^\mu-e A^\mu,
\end{equation}
with $k_Y$ and $k_{\bar{Y}}$ the four-momentum of the hyperon and anti-hyperon in the vertex $\Gamma_{\psip Y\bar{Y}}$, respectively. Adding the generated four-point $\psip Y\bar{Y}\gamma$ contact terms into the A and B-type hadronic operators of the octet hyperons, one can get the gauge invariant elastic-pole contributions to the amplitude of $\psip \to Y\bar{Y}e^+e^-$. The contact term $H_{\text{cont.}}^{Y,\mu}$ corresponding to the vertex $\Gamma_{\psip Y\bar{Y}}$ of Eq.~\eqref{eq:JpsiN} is given by
\begin{align}
    H_{\text{cont.}}^{Y,\mu}=&\,\bar{u}_{s_p}(p_3)v_{s_{\bar{p}}}(p_4)\Bigg\{\frac{2 e}{3M_{\psip}^2(M_{\psip}+2m_Y)}\notag\\
    &\phantom{xxx}\times\bigg(4 M_{\psip}^2 g_D (p_3-p_4)\cdot \epsilon_{s_{\psip}} (p_4^\mu-p_3^\mu)+4 g_D (p_3-p_4)\cdot \epsilon_{s_{\psip}} (p_3-p_4)\cdot p_0 p_0^\mu\notag\\
    &\phantom{xxxxxx}+\left(4g_D\left((p_3 \cdot p_0)^2+(p_4 \cdot p_0)^2\right)+M_{\psip}^2(2m_Y g_D (3M_{\psip}+2m_Y)-3 g_S)\right)\epsilon_{s_{\psip}}^\mu\bigg)\Bigg\}\notag\\
    &+\bar{u}_{s_p}(p_3)\slashed{\epsilon}_{s_{\psip}} v_{s_{\bar{p}}}(p_4) \frac{4 e g_D}{3 M_{\psip}^2}\left((p_3-p_4)\cdot p_0 p_0^\mu+M_{\psip}^2(p_4^\mu-p_3^\mu) \right)\notag\\
    &+\bar{u}_{s_p}(p_3)\slashed{p}_0 v_{s_{\bar{p}}}(p_4) \frac{-2 e g_D}{M_{\psip}^2}\left( (p_3-p_4)\cdot \epsilon_{s_{\psip}} p_0^\mu+(p_3-p_4)\cdot p_0 \epsilon_{s_{\psip}}^\mu\right)\notag\\
    &+\bar{u}_{s_p}(p_3)\gamma^\mu v_{s_{\bar{p}}}(p_4) \left(2 e (p_3-p_4)\cdot \epsilon_{s_{\psip}}\right).
\end{align}
It should also be mentioned that the contact terms obtained by the substitution treatment can only restore the gauge invariance exactly when the Dirac form factor of $\Sigma^+$ equals 1, i.e., $F_1(q^2)=1$ ($F_1(q^2)=-1$ for $\Xi^-$). This does work in the case where the photon is real. For the virtual photon involved processes, however, the off-shell effects of the vertex $\gamma \Sigma^+\bar{\Sigma}^-$ have to be taken into account to preserve gauge invariance strictly~\cite{Scherer:1996ux,Haberzettl:2006bn}. A phenomenological off-shell vertex function proposed by Ref.~\cite{Gross:1987bu} is used here for simplicity, since such off-shell contribution is not uniquely defined. For $\Sigma^+$, it reads
\begin{equation}
    \Gamma_{\text{off}}^\mu(q)=i e \left(\gamma^\mu F_1(q^2)+\frac{1-F_1(q^2)}{q^2}q^\mu\slashed{q}+\frac{i\sigma^{\mu\nu}}{2m_{\Sigma^+}}q_\nu F_2(q^2)\right),
\end{equation}
and similarly for $\Xi^-$.
Combining the contact terms $H_{\text{cont.}}^{Y,\mu}$ and the off-shell vertex function $\Gamma_{\text{off}}$, one can easily show that the gauge invariance is preserved. Note that whether or not considering the $\gamma YY$ off-shell effects, the difference is negligible in our numerical results. Thus, the off-shell contributions of the $\gamma YY$ vertex are dropped in the present work.

\subsection{C. Application to the proton}
In this subsection, the radius extraction from the Dalitz decay of $\jpsi \to p\bar{p}e^+e^-$ for the proton is investigated. Figure~\ref{fig:proton} presents the differential ratio $R$, as defined in Eq.~\eqref{eq:raND} with the $\Delta(1232)$ as the pertinent decuplet state, and the sensitivity of the $e^+e^-$ invariant-mass distribution to the proton charge radius $r^p_E$ for the simulation of the \bes{} experiment. This result shows that 
for $m_{p\bar{p}}\gtrsim2.7\,{\rm GeV}$, the type-A and type-B decays can be accurately described by the proton pole contribution to the $\jpsi\to p\bar{p}\gamma_{\rm FSR}^*$ process at the percent level, which is consistent with the fact that there are no any visible $m_{p\gamma}$ or $m_{\bar p\gamma }$ bands in the Dalitz plot shown in Ref.~\cite{Kappert:2022fox}. The expected event yields for the radius extraction of the proton from the $\jpsi$ data sets at \bes{} and STCF are estimated to be approximately $3\times 10^3$ and $1\times 10^6$, respectively. Fitting the Monte Carlo generated data with 0.84~fm as the input value leads to $r_p^E=0.71(9)$~fm  and $0.845(4)$~fm for the simulations of the \bes{} and STCF experiments, respectively. As shown in the right panel, varying the amplitude parameters (such as the $S/D$ ratio $R_{S/D}$ of the $J/\psi p \bar p$ vertex within the range of the ratio for similar decays obtained in Ref.~\cite{Wu:2021yfv}, $g_X$ and $\Lambda_X$) have a negligible effect on the accuracy of radius extraction at the current \bes{} experimental setup (the gray band completely lies within the $1\sigma$ region of the best solution). While for the STCF experiment, including the three amplitude parameters ($R_{S/D}$, $g_X$ and $\Lambda_X$) as additional degrees of freedom, a combined fit reveals an additional 0.5\% uncertainty arising from those model parameters, leading to $r_p^E=0.845(7)$~fm. 
The sensitivity analysis implies that a high precision can be achieved in extracting the proton charge radius using the proposed novel approach. This method offers an independent cross-check on the proton radius extracted from elastic $ep$ scattering data or from spectrocopy of leptonic hydrogen.

\begin{figure}[tb]
	\centering
	\includegraphics*[width=0.48\textwidth,angle=0,scale=1.0]{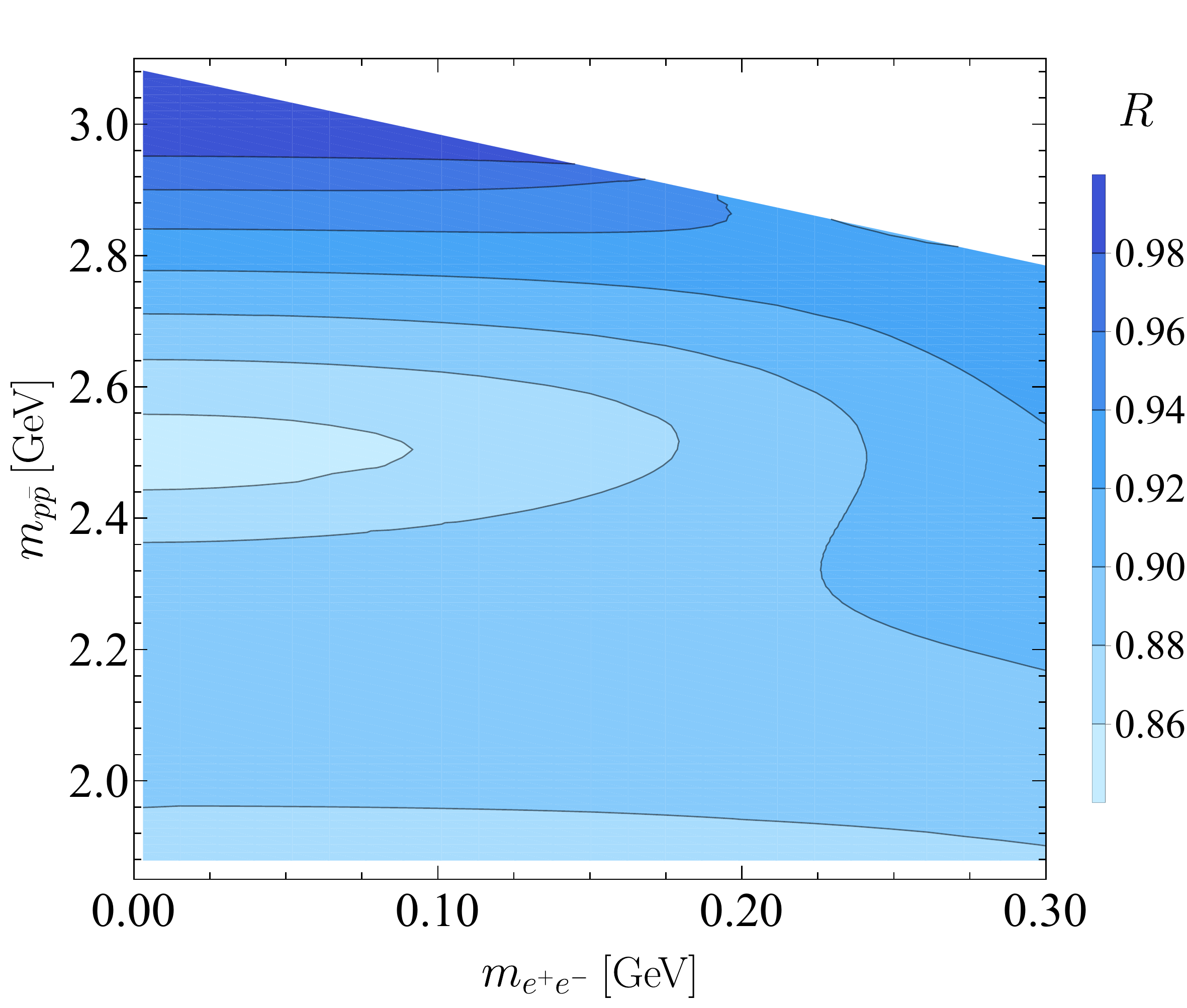}\qquad
	\includegraphics*[width=0.47\textwidth,angle=0]{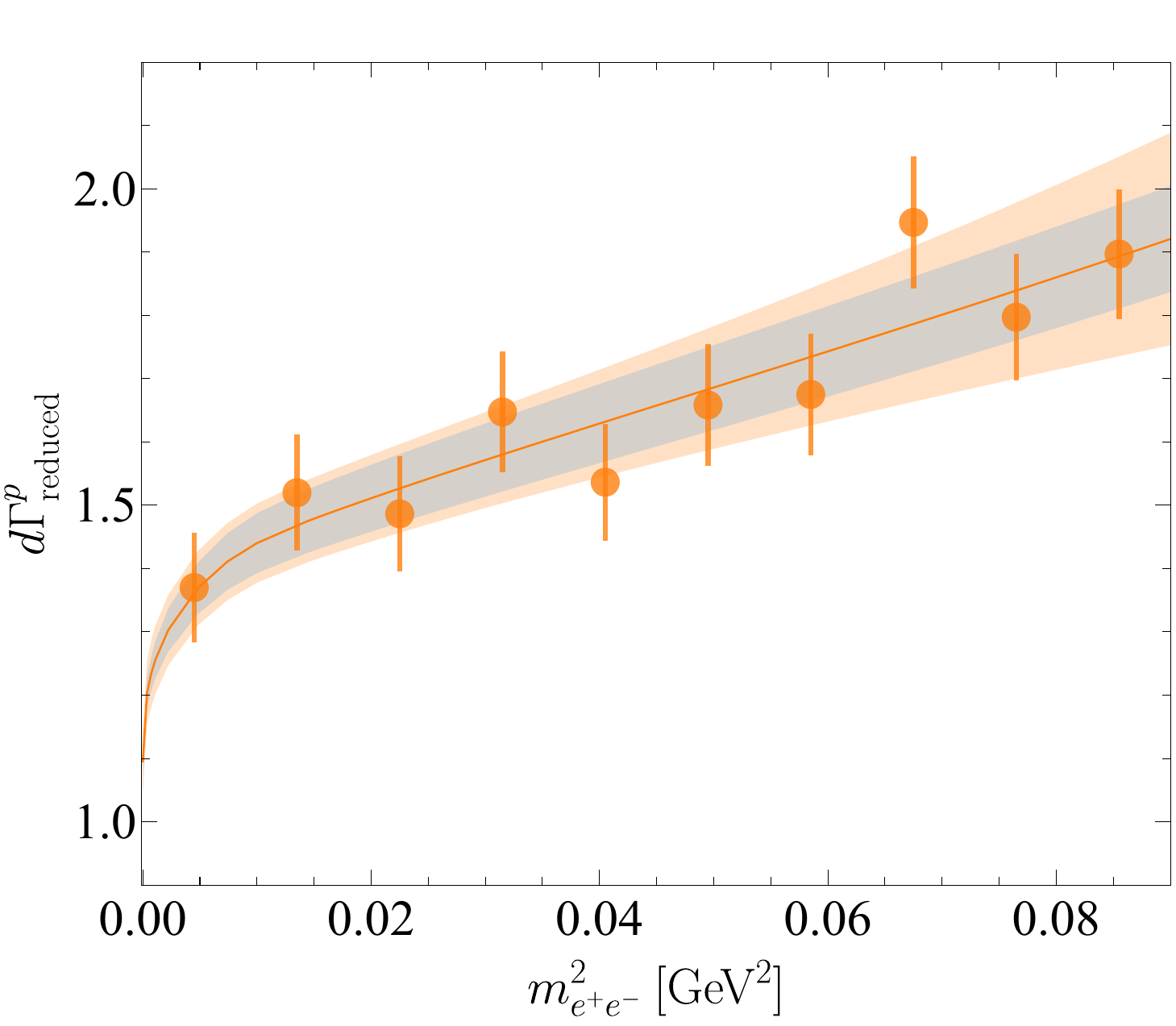}
	\caption{
		The left panel presents the differential cross section ratio $R$ defined in Eq.~\eqref{eq:raND} for the proton. The right panel shows the fit to the synthetic data of $3\times 10^3$ proton Monte Carlo events. The extracted proton charge radius from the data set fitting is $(0.71\pm 0.09)$~fm. And the gray band shows the effects of model parameters to the radii extractions, as detailed in the text. 
	\label{fig:proton}}
\end{figure}

\end{onecolumngrid}
\end{appendix}

\end{document}